\pdfoutput=1
\documentclass[iop,twocolumn,apjl,letterpaper]{emulateapj}

\usepackage{graphicx}
\usepackage{amsmath}
\newcommand       \rhocgs       {\,{\rm g~cm}^{-3}}

\begin{document}
\title{Breeding Super-Earths and Birthing Super-Puffs in Transitional Disks}

\author{Eve J. Lee\altaffilmark{1}, Eugene Chiang\altaffilmark{1,2}}
\altaffiltext{1}{Department of Astronomy, University of California Berkeley, Berkeley, CA 94720-3411, USA; evelee@berkeley.edu, echiang@astro.berkeley.edu}
\altaffiltext{2}{Department of Earth and Planetary Science, University of California Berkeley, Berkeley, CA 94720-4767, USA}

\begin{abstract}
The riddle posed by super-Earths (1--4$R_\oplus$,
2--20$M_\oplus$) is that they are not Jupiters:
their core masses are large enough to trigger
runaway gas accretion, yet somehow super-Earths  
accreted atmospheres that weigh only
a few percent of their total mass.
We show that this puzzle is solved if
super-Earths formed late, as the last vestiges
of their parent gas disks were about to clear. This scenario
would seem to present fine-tuning problems, but we
show that there are none. Ambient gas
densities can span many (in one case up to 9) orders of magnitude,
and super-Earths can still robustly emerge 
after $\sim$0.1--1 Myr with
percent-by-weight atmospheres.
Super-Earth cores are naturally bred in gas-poor
environments where gas dynamical friction has
weakened sufficiently to allow constituent
protocores to gravitationally stir one another
and merge. So little gas is present
at the time of core assembly that cores
hardly migrate by disk torques: formation of super-Earths
can be in situ. 
The basic picture --- that close-in super-Earths form
in a gas-poor (but not gas-empty) inner disk,
fed continuously by gas that bleeds inward from a more massive
outer disk ---
recalls the largely evacuated but still accreting 
inner cavities of transitional protoplanetary disks.
We also address the inverse
problem presented by super-puffs: an uncommon class of
short-period planets seemingly too voluminous for their small
masses (4--10$R_\oplus$, 2--6$M_\oplus$).
Super-puffs most easily acquire their
thick atmospheres as dust-free, rapidly cooling worlds outside
$\sim$1 AU where nebular gas is colder, less dense,
and therefore less opaque.
Unlike super-Earths which can form in situ, super-puffs
probably migrated in to their current orbits; they
are expected to form the outer links of mean-motion
resonant chains, and to exhibit greater water
content. We close 
by confronting observations and itemizing
remaining questions.
\end{abstract}

\section{Introduction}
\label{sec:intro}

The {\it Kepler} spacecraft has established 
that about half of all Sun-like stars 
harbor at least one planet \citep[e.g.,][]{fressin13}. 
Of these, the most common are 
``super-Earths,'' here defined
as those objects having radii 
between 1 and 4 $R_\oplus$.
Super-Earths are found orbiting
$\sim$60\% of FGK-type dwarfs with
periods $< 85$ days
\citep{howard10,batalha13,petigura13,dong13,fressin13,rowe14}.
Transit-timing analyses \citep{wu13} and 
radial velocity surveys \citep{weiss14} 
have determined that the typical masses of super-Earths 
are $\sim$2--20$M_\oplus$.
These masses overlap with the range of core masses believed to
trigger the formation of gas giants,
according to the theory of 
core-nucleated instability 
(e.g., \citealt{mizuno80}, \citealt{stevenson82}, 
\citealt{pollack96}, \citealt{ikoma00}).
Yet super-Earths are not Jupiters; the
gas-to-core mass ratios (GCRs) of super-Earths
are of order 1--10\% \citep[e.g.,][]{lopez14, 
rogers10-general, rogers10-gj1214b}.
They are too small to be gas giants
and too large to be purely rocky.\footnote{Structurally
they most resemble Neptune ($4 R_\oplus$, $17 M_\oplus$,
GCR $\simeq 10$\%),
prompting some to call them ``mini-Neptunes'' or ``sub-Neptunes''.
We will stick to ``super-Earths''.}

Motivated by these discoveries,
we have investigated in a series of papers
the physics of nebular accretion: how rocky cores at stellocentric
distances of 0.1--5 AU acquire gas from
their parent nebulae (\citealt{paper1}, 
Paper I; \citealt{paper2}, Paper II). 
The rate at which planets siphon gas from their
natal disks is limited by 
how fast such gas can cool.
As atmospheres cool, they shrink,
allowing nebular gas to refill the planets' Hill (or Bondi) spheres.
Atmospheric cooling is regulated by the thermodynamic 
properties of radiative-convective boundaries (rcb's).
In dusty atmospheres, rcb properties
are set by the microphysics of dust sublimation
and H$_2$ dissociation, not by external nebular conditions.
Regardless of their orbital distance,
dusty atmospheres atop cores of mass $\gtrsim 5 M_\oplus$
undergo runaway gas accretion to spawn Jupiters, when
embedded in gas-rich disks (Paper I).\footnote{ ``Runaway'' 
ensues when the atmospheric mass becomes
comparable to the core mass: when the self-gravity of
the gas envelope strongly enhances the cooling rate. 
Runaway is a thermodynamic catastrophe,
not a dynamic one; the envelope remains
hydrostatic throughout. To maintain
hydrostatic equilibrium as the envelope's
self-gravity grows, the cooling luminosity rises dramatically,
hastening the cooling and contraction of the envelope
and thereby causing the Hill (or Bondi) sphere
to refill at a faster rate. This positive feedback loop between
self-gravity, cooling, and accretion causes the planet
to balloon into a gas giant.}
Dust-free atmospheres have rcb's that do depend
on nebular environment;
nevertheless, because they generally cool faster than
dusty atmospheres, they are also prone to gravitational
collapse essentially everywhere in protoplanetary nebulae (Paper II). 
The threat of runaway
is all the more serious because it cannot be neutralized by heating from
infalling planetesimals; such heating tends to be overwhelmed by
the acceleration of gas accretion from increasing
core mass (Paper II).

Paper I proposed two ways to circumvent runaway.
In scenario (a), super-Earths can stay as super-Earths,
even in gas-rich nebulae, provided their atmospheres
are chock full of dust. Supersolar dust-to-gas ratios
render atmospheres so opaque that their cooling times
exceed the nebular lifetime.
This scenario finds empirical support in the
supersolar metallicities of observed super-Earth atmospheres  \citep[e.g.,][]{kreidberg14,knutson14,morley13}.
In scenario (b), super-Earth cores did not nucleate
Jupiters simply because they were born at the tail end
of the gas disk's life, with only dregs of gas
remaining for $\sim$1 Myr or so.
The formation of cores from smaller mass protocores
is arguably delayed until disk gas is largely
depleted: only then is gas dynamical friction
weak enough to permit protocores to gravitationally
stir each other, cross orbits, and merge (see, e.g., \citealt{dawson15}).

In the present paper, we revisit these two proposed formation
pathways for super-Earth atmospheres.
In Sections 2 and 3, respectively,
we re-examine scenarios (a) and (b),
assessing more critically their strengths and weaknesses
to decide whether a gas-rich or gas-poor formation
environment is more likely. A concern common to 
both scenarios is that fine-tuning of parameters
is needed to obtain desired outcomes. We will
confront this concern head-on.

The conundrum of super-Earths is that many of them
have too little gas for their large core masses. 
The flip side of this puzzle is presented
by ``super-puffs'': a rare class of {\it Kepler}
planets characterized by orbital periods $\lesssim 50$ days,
radii $\sim$4--10$R_\oplus$, masses $\sim$2--6$M_\oplus$, 
and inferred GCRs $\gtrsim$ 20\% \citep[e.g.,][]{masuda14,lopez14}.
Super-puffs seem to have too much gas for their small core masses. 
To attain their large GCRs, super-puff cores must have enjoyed
an environment that enabled rapid atmospheric cooling.
In Section 4, we show how such an environment
obtains at large stellocentric distances, 
outside $\sim$1 AU --- 
thus implicating orbital migration for the origin
of super-puffs.

Our series of three papers studies how planetary atmospheres
are built from nebular accretion, with the ultimate
goal of explaining how super-Earths and super-puffs
got their gas. The solutions we provide 
inspire new questions and point to future areas of improvement,
as well as ways in which our ideas can be tested by observers.
We attempt to provide this ``big-picture'' view in Section 5,
where we summarize our results and chart the way forward.

All calculations of nebular accretion
throughout this paper derive from the numerical
models of passively cooling atmospheres of Paper I, or their
analytic counterparts in Paper II. Readers interested
in the underlying machinery 
should consult those studies.
We have tried to write this third paper so that it can be understood
with as few technical references as possible,
relegating such information mostly to figure captions.

\section{Dangers of Runaway \\in Gas-Rich, Metal-Rich Nebulae (Scenario A)}
\label{sec:sceA}

Increasing the dust content --- and by extension the metallicity ---
of nebular gas increases the opacity of the accreted atmosphere.
Dustier atmospheres cool more slowly, accrete less
efficiently, and are thus less prone to runaway (e.g.,
\citealt{stevenson82}; \citealt{ikoma00}; \citealt{piso14};
Paper I; \citealt{piso15}).
High metallicities for super-Earth atmospheres
are also motivated by observations: the notoriously
flat transmission spectra of GJ 3470b, GJ 1214b, 
and GJ 436b can be attributed to clouds or hazes that  
form only in environments 
with supersolar metallicities (e.g., $\sim$50 $\times$ solar;
\citealt{morley13,crossfield13,biddle14,kreidberg14,knutson14}).
The atmosphere of HAT-P-11b may have patchy clouds
(Line \& Parmentier, in preparation)
or be cloud-free;
the latter interpretation still implicates a
supersolar metallicity ($Z \gtrsim 0.6$),
as judged from the planet's water absorption feature \citep[][their Figures 2 and 3]{fraine14}.

If the effects of raising the dust-to-gas ratio were limited
to increasing the opacity, then high
atmospheric metallicities could explain
how super-Earths avoided runaway even in gas-rich nebulae.
Life, unfortunately, is not so simple.
We can identify two reasons why
dust/metal enrichment may not save super-Earths
from exploding into Jupiters, in disks with
full complements of gas:
\begin{enumerate}
\item High metallicity gas has high mean
molecular weight and is therefore
more susceptible to gravitational collapse
(e.g., \citealt{hori11}; \citealt{nettelmann11}; 
Paper II).
\item Grains can coagulate
and sediment out of radiative layers, forfeiting
their contributions to the opacity
(e.g., Mordasini 2014; Ormel 2014).
\end{enumerate}
We quantify these weaknesses below.

\vspace{0.05in}

Throughout this paper, our proxy for a gas-rich
disk is the ``minimum-mass extrasolar nebula'' (MMEN)
of \citet{chiang13}: a solar-composition
disk having enough solid material to form {\it Kepler}-like
systems. Its gas surface
density is $\Sigma_{\rm MMEN} = 4 \times 10^5 \,(a/0.1\,{\rm AU})^{-1.6}$ g/cm$^2$, which is about 5$\times$ larger than
that of the conventional minimum-mass solar nebula.
The assumed temperature profile
is $T_{\rm MMEN} = 1000 \,(a/0.1\,{\rm AU})^{-3/7}$ K, and the volumetric
gas density is $\rho_{\rm MMEN} = 6 \times 10^{-6} \, (a/0.1\,{\rm AU})^{-2.9}$ g/cm$^3$.
``Dusty'' models assume
a relative grain size distribution similar to that of the
diffuse interstellar medium \citep{ferguson05},
while ``dust-free'' models take all metals to be in the gas
phase at their full assumed abundances.

\begin{figure}[!tbh]
\centering
\includegraphics[width=0.5\textwidth]{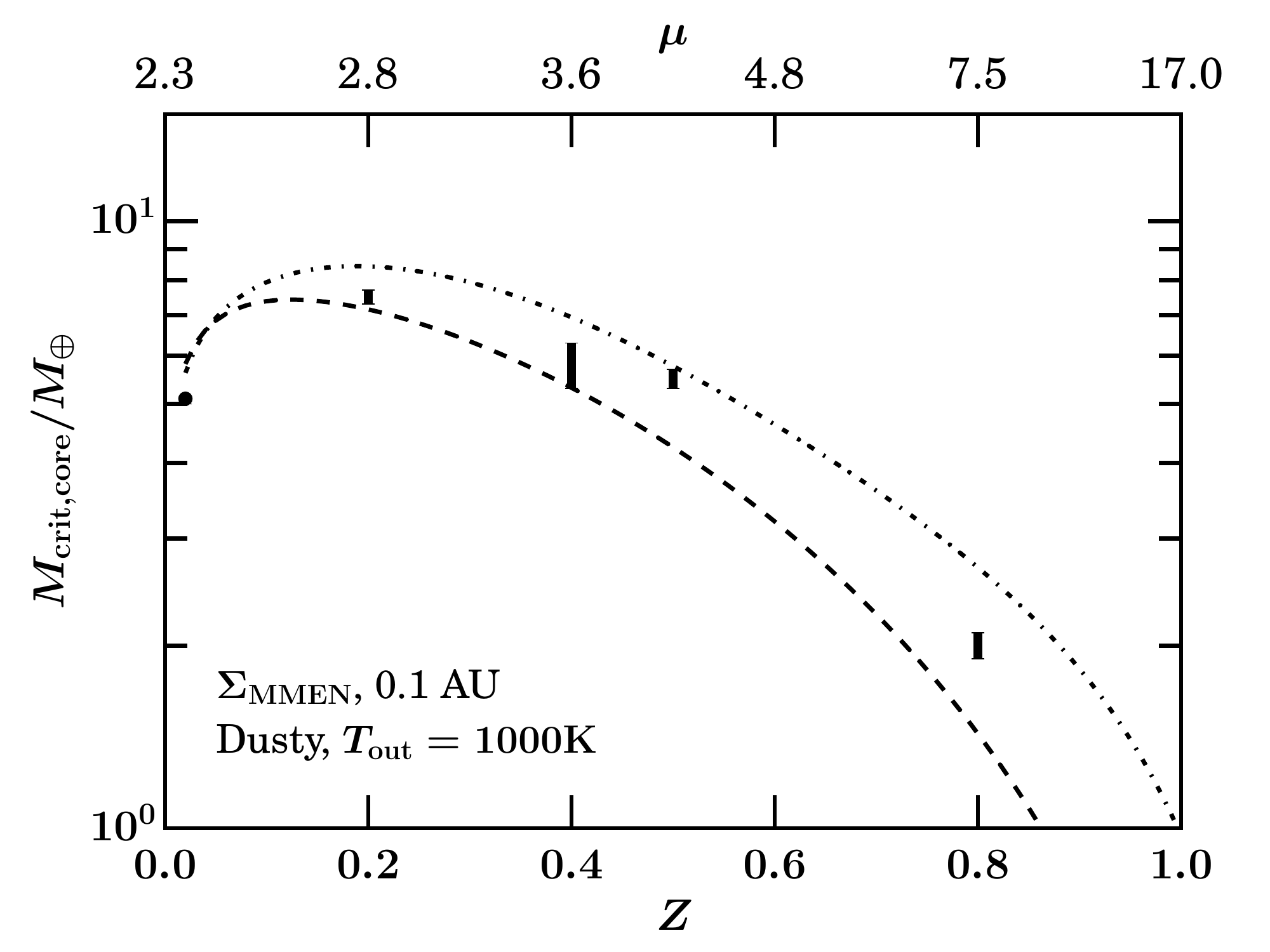}
\caption{\label{fig:Mcrit_Z}
Critical core masses 
$M_{\rm crit, core}$ do not rise monotonically with
dust-to-gas ratio $Z$. For $Z \gtrsim 0.2$, atmospheres are actually
more susceptible to runaway because of the increasing molecular
weight $\mu$ (top axis).
Results are calculated for dusty atmospheres
at 0.1 AU in the ``gas-rich" MMEN.
The lower limit on each error bar is the maximum $M_{\rm core}$ 
that can be numerically evolved;
for $Z=0.02$--0.8, their runaway times (when planetary luminosities
reach their minima and start increasing) are 10.5, 11.7, 
17.8, 12.7, and 14 Myr, in order of increasing plotted $Z$.
Each upper limit is derived by scaling the
corresponding lower limit to a runaway time of 10 Myr 
(our assumed lifetime for a gas-rich disk) using the
empirical relation
$t_{\rm run} \propto M_{\rm core}^{-3.93}$ (Paper I).
Analytic predictions for $M_{\rm crit,core}$ (Paper II)
are overplotted as curves. The dashed curve is computed using 
gas-to-core mass ratio
GCR = 0.5, time $t = 10$ Myr, the dimensionless
adiabatic gradient $\nabla_{\rm ad} = 0.17$, and
the radiative-convective boundary temperature
$T_{\rm rcb}=2500$ K. The dash-dotted curve
is the same except that it uses $T_{\rm rcb}$
as computed from the numerical model for every black datum.
}
\end{figure}

\subsection{High-$\mu$ Catastrophe}
\label{ssec:highZ}

Metal enrichment breeds atmospheres with high mean molecular weight $\mu$.
Such atmospheres more easily collapse under their own weight, triggering
the formation of gas giants.
Figure \ref{fig:Mcrit_Z} illustrates the danger of having too
high a metallicity $Z$. As $Z$ increases from 0.02 to 0.2, the critical core mass
$M_{\rm crit,core}$ for runaway gas accretion --- evaluated at 0.1 AU in
a dusty MMEN --- increases
from 5 to 8$M_\oplus$
as rising opacities impede cooling and accretion.
But further gains
in $M_{\rm crit,core}$ with $Z$ are not to be had;
for $Z \gtrsim 0.2$,
enhancements in opacity are outweighed by increases in
$\mu$. The critical core mass stays at about 5--6$M_\oplus$ for $Z = 0.5$--0.6 ($\mu \simeq 4$),
and drops to as low as $2 M_\oplus$ for $Z = 0.8$ $(\mu \simeq 7.5)$.\footnote{Critical core masses can
decrease even further because increasing $Z$
increases the potential
for molecular chemistry and phase changes,
both of which consume energy and render gas more
isothermal, steepening density gradients.
These effects are more significant at larger orbital
distances where gas is colder and 
hosts more molecules;
see Section \ref{sec:superpuff} for more details.}
Here we define the critical core mass
such that its runaway time (when the atmospheric cooling luminosity
attains a minimum; see Papers I and II) equals 
the total disk lifetime of $\sim$10 Myr 
\citep{mamajek09,alexander14}.

The similarity of the critical core masses plotted
in Figure \ref{fig:Mcrit_Z} to observed super-Earth
masses \citep[e.g.,][]{weiss14} is a primary
reason for believing that super-Earths did not form
in a gas-rich disk. At no metallicity is a
10$M_\oplus$ core embedded in an undepleted
nebula able to remain a super-Earth.
We shall see in the next section that
even the modest advantage presented by
a 10--20$\times$ dust-enhanced disk 
in staving off runaway may be erased by grain growth.

\subsection{Dust Coagulation: Low-$\kappa$ Catastrophe}
\label{ssec:dust-free}

Opacity gains by dust depend on the grain size
distribution and the degree to which dust is
well mixed in gas. Dust coagulation can wipe out
these gains, not only 
by reducing the surface area
of dust per unit mass (e.g.,
\citealt{piso15}), but also by speeding
the gravitational sedimentation of dust 
out of the atmosphere's radiative layers 
\citep{mordasini14,ormel14}.

Grain growth in disks 
can be fast, especially 
at the high densities 
that prevail inside 1 AU 
\citep[e.g.,][]{blum00, windmark12}.
Grains need only grow to sizes 
$\gtrsim 0.1$ mm for the 
gas opacity to overwhelm 
the dust opacity (in the Rosseland mean sense).
Dust can also be removed 
by drifting to deeper, hotter 
layers where they sublimate.
This last concern applies 
to atmospheres that are 
purely radiative (assumed static) 
from the outermost surface layers down
to the sublimation front.
For our models at $\sim$0.1 AU, 
such conditions obtain for gas-to-core mass
ratios (GCRs) $>$ 0.1. 
We estimate settling times in such radiative
layers to be 
$\sim$1 Myr --- shorter than
disk lifetimes of $\sim$5--10 Myr --- for
$\mu$m-sized grains.
Larger grains settle even faster.

\begin{figure}[!tbh]
\centering
\includegraphics[width=0.5\textwidth]{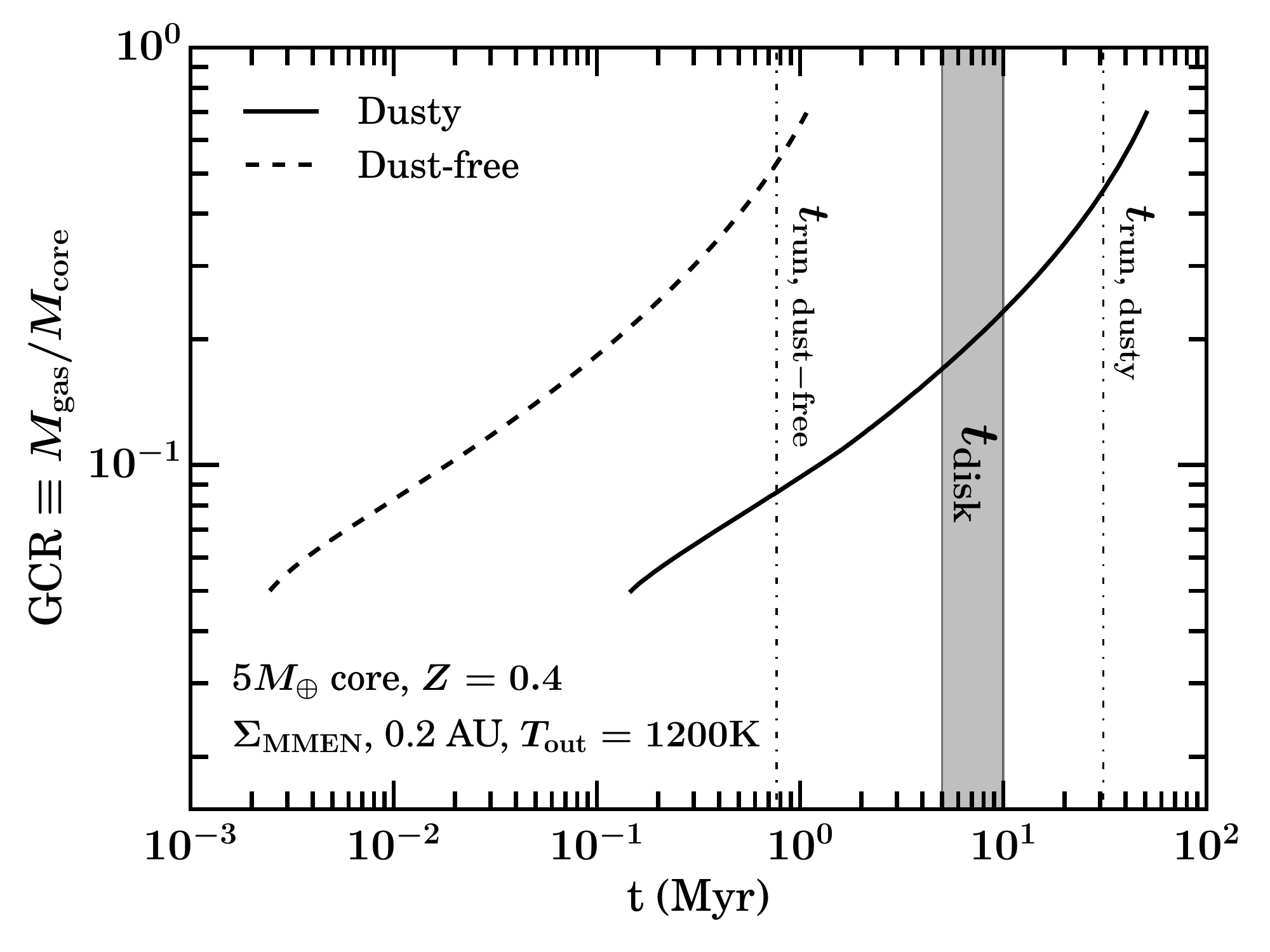}
\caption{\label{fig:dust-free}
A tale of two opacities:
evolution of gas-to-core 
mass ratios for a 5$M_\oplus$ 
core situated at 0.2 AU in a metal-enhanced 
($Z=0.4$) MMEN with either ``dusty" or ``dust-free" opacities.
``Dusty" opacities assume a grain size distribution
similar to that of the interstellar medium \citep{ferguson05};
"dust-free" opacities eliminate the contribution from grains,
simulating their removal by coagulation and settling 
\citep[e.g.,][]{ormel14}.
For Figure \ref{fig:dust-free}
and Figure \ref{fig:dust-free-comp},
the nebular temperature $T_{\rm out}$ was enhanced
by a factor of 1.6 above the MMEN value to 1200 K;
this enhancement (which is still
physically plausible; see, e.g., \citealt{dalessio01})
was necessary to numerically resolve the outer radiative zone.
Runaway times mark when model cooling luminosities
reach their minima (see Papers I and II) and are indicated
by vertical dot-dashed lines.
Dust-free atmospheres run away well 
before the disk dissipation time 
$t_{\rm disk}\sim 5$--10 Myr.
}
\end{figure}

We demonstrate in Figure \ref{fig:dust-free} 
the drastic effect of rapid dust coagulation 
and sedimentation: 
even at $Z=0.4$, 
5$M_\oplus$ cores with no dust in their atmospheres 
run away within the disk lifetime. 

\begin{figure}[!tbh]
\centering
\includegraphics[width=0.5\textwidth]{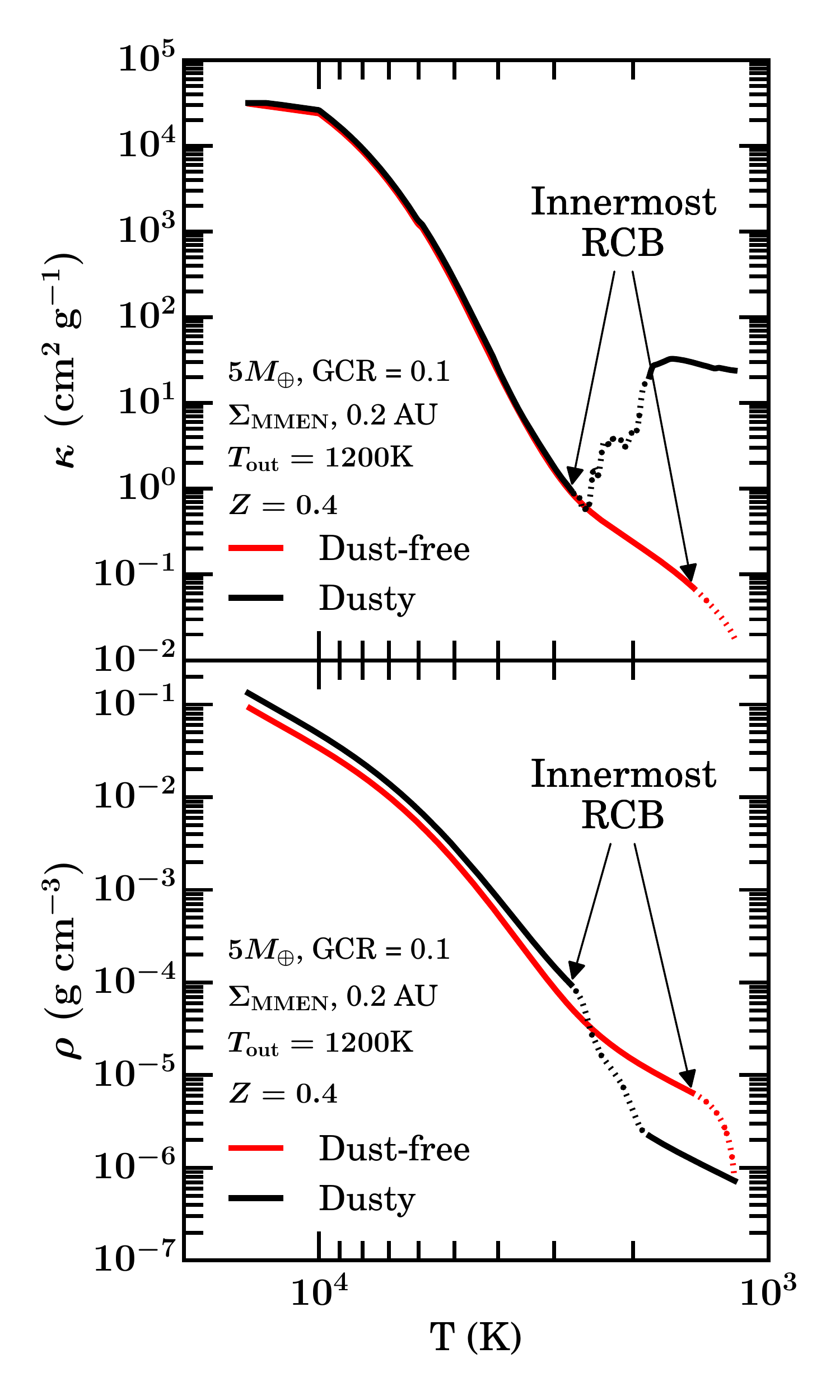}
\caption{\label{fig:dust-free-comp}
The effect of dust on the radiative-convective boundary (rcb).
Top and bottom panels show opacity and density profiles of
the 5$M_\oplus$ models presented in
Figure~\ref{fig:dust-free} 
(nebular temperature 
enhanced by a factor of 1.6 at 0.2 AU to 1200 K), 
evaluated at GCR=0.1,
for dusty vs.~dust-free opacities (the latter is our proxy for 
grain growth and sedimentation). Dashed lines delineate radiative 
zones while solid lines trace convective zones.
Dust sublimation in a dusty atmosphere
forces $\kappa$ to drop by approximately 2 orders of magnitude,
opens a radiative window between 2000 and 2700 K,
and pushes the innermost rcb (the one that controls the overall
luminosity of the atmosphere) to greater depth.
By comparison, the rcb in a dust-free atmosphere
is situated at higher altitude where
$\kappa$, $\rho$, and $T$ are lower
by factors of 14, 14, and 2, respectively;
the upshot is that
$L \propto T^3/\rho\kappa$ 
increases by a factor of 30. The higher $L$ explains 
the shorter $t_{\rm run,dust-free}$ in Figure~\ref{fig:dust-free}.}
\end{figure}

Dust-free atmospheres accrete
faster than dusty atmospheres because 
removing the opacity from dust lifts the
radiative-convective boundary (rcb)
to higher altitudes where atmospheres cool faster
(Paper II).
The reason for the higher altitude of the rcb is 
as follows.
In dusty atmospheres,
a radiative window is necessarily opened
in the planet's deep interior where
temperatures are high enough ($T \gtrsim 2000$ K)
for dust to sublimate, dropping the opacity $\kappa$ 
by 2 orders of magnitude and facilitating
energy transport by radiation.
This radiative window closes at the H$_2$ sublimation
front at $T \sim 2500$ K; here lies
the innermost rcb, where $\kappa$ surges back up from
the creation of H$^-$ ions.
Dust-free atmospheres have no such radiative window
because there is no dust to sublimate.
Their rcb's are located
near their outermost boundaries, at the base
of a nearly isothermal radiative layer
that connects directly to the ambient disk.
Figure \ref{fig:dust-free-comp}
reveals the rcb in a dust-free atmosphere to
occur at $T\sim 1500$ K, close to the nebular
temperature $T_{\rm out} = 1200$ K.
The dust-free rcb is located
at a higher altitude --- where density $\rho$ and opacity $\kappa$
are lower --- as compared to the innermost
rcb in a dusty atmosphere. 
Because an envelope's cooling
luminosity is controlled at its rcb
where $L \propto 1/(\rho \kappa)$ (e.g., equation 33 in Paper I),
the lower $\rho$
and $\kappa$ characterizing the rcb in a dust-free
atmosphere leads to larger $L$, faster cooling,
and more rapid accretion of gas (see also Paper II).

\section{Acquiring Atmospheres in Gas-Poor Nebulae \\(Scenario B)}
\label{sec:sceB}

A way out of the high-$\mu$
and low-$\kappa$ catastrophes 
(Sections \ref{ssec:highZ}--\ref{ssec:dust-free})
is to form super-Earths in nebulae
depleted in gas. 
Less ambient gas would obviously seem to reduce
the likelihood of runaway. It does, but
final GCRs are remarkably insensitive to ambient
nebular densities, as we will stress later in this section.
Just as important, if not more so,
is the limited time
available for gas accretion in scenario (b):
staging envelope accretion during
the era of disk dispersal sets 
a stricter time limit on how much gas a core can accrete.
Observations of transitional protoplanetary disks 
suggest that their inner regions clear
in $\lesssim 10\%$ 
of the disk's total lifetime \citep{alexander14}.
If close-in cores start accreting gas 
during this final clearing phase, 
then they have $\lesssim 1$ Myr to build
their atmospheres rather than $\sim$10 Myr;
they run out of time before running away.

In addition to the advantages presented
by depleted, short-lived nebulae in avoiding
gas giant formation,
there is actually a physical
reason to believe super-Earth cores
of mass $\sim$2--20$M_\oplus$ 
form in gas-poor environments.
Cores assemble from the mergers 
of smaller ``protocores," whose eccentricities
need to grow large enough for their orbits
to cross. Mutual gravitational scatterings between
protocores can excite eccentricities, 
but must compete against eccentricity damping
by tidal torques (a.k.a.~dynamical friction)
exerted by gas.
Core formation is therefore 
delayed until the gas disk depletes
sufficiently that eccentricity damping
no longer prevents mergers.\footnote{This argument
is subject to the possibility that gas
might actually excite eccentricities,
either resonantly by Lindblad torques
\citep{goldreich03,duffell15}
or stochastically by turbulent density fluctuations
\citep[e.g.,][and references therein]{ogihara07}.}

In Section \ref{ssec:gas-depletion-factors}, 
we estimate the extent to which nebular gas densities
can be reduced and still give rise to the GCRs
inferred for super-Earths. 
There we also connect our 
idea that super-Earths form in gas-poor disks
to theories of transitional disks
and disk dispersal. Whether the gas depletion factors we require
are compatible with protocore mergers and core assembly
is examined in Section \ref{ssec:deplete_merge}.

\subsection{Gas Depletion Factors Compatible with
GCRs and Disk Dispersal Theories}
\label{ssec:gas-depletion-factors}

One might think that for a planet to achieve
a GCR $\gtrsim 1$\%, it needs to form in
a disk whose gas-to-solid ratio $\gtrsim 1$\%.
Such reasoning is too restrictive; the argument presumes that
once a rocky core consumes all the gas in its immediate
vicinity, it cannot accrete more because
its local gas reservoir is not replenished.
But replenishment is the hallmark of accretion
disks: the inner disk, where cores reside,
can be resupplied with fresh gas by diffusion from the outer disk.
In principle, a close-in core can
sit in an environment nearly devoid of gas,
yet still acquire an atmospheric
mass fraction $\gtrsim 1$\%, provided
gas is fed to it from the outside for a long enough time 
(i.e., over the $\sim$Myr timescales that
our calculations cover; the assumption
of all our models is that gas densities
at the planet's outer boundary --- the Hill
or Bondi radius --- are fixed in time).

\begin{figure*}[!tbh]
\centering
\includegraphics[width=\textwidth]{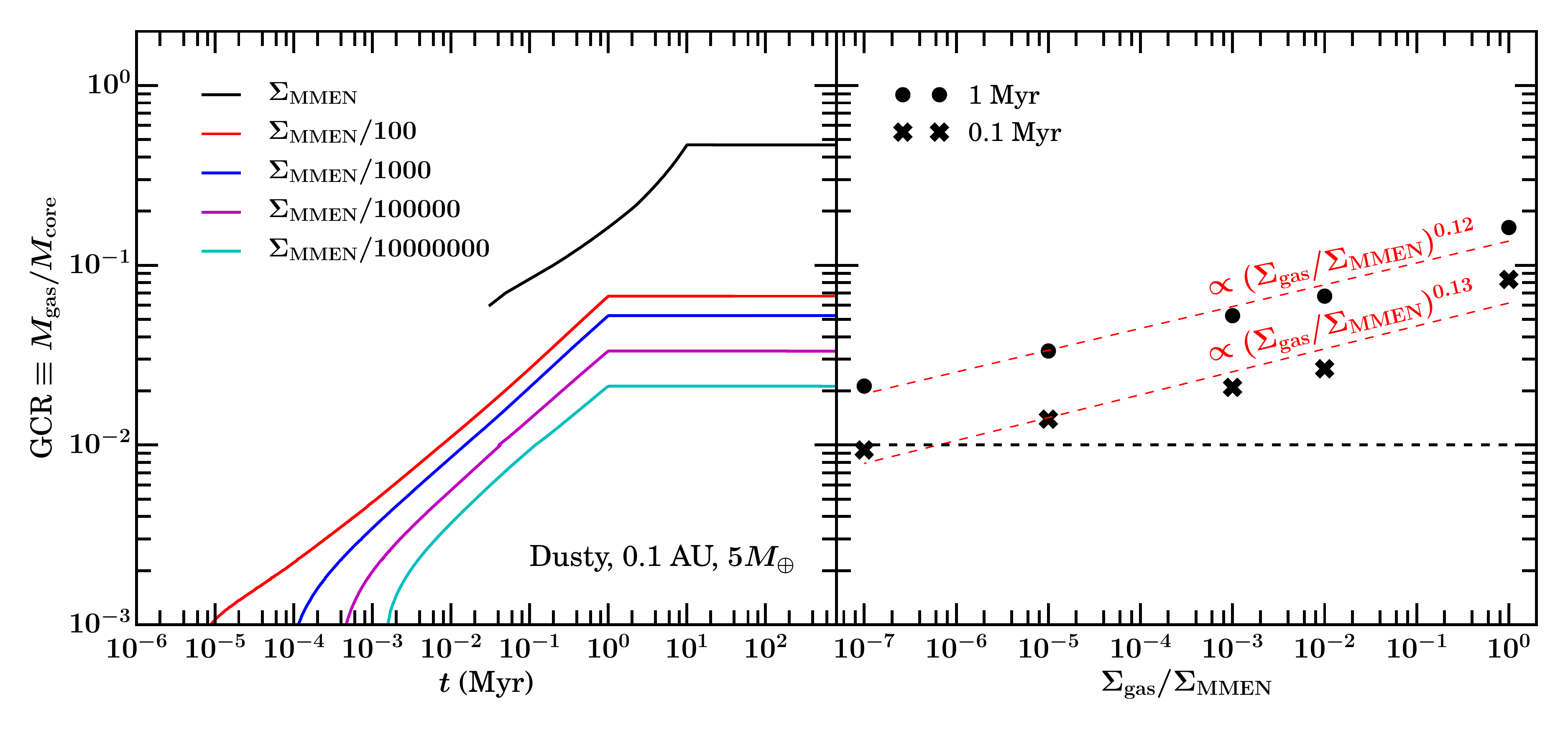}
\caption{
Under scenario (b), there is an impressive variety of short-lived,
gas-poor nebulae that can generate GCRs on the order of a few
percent, similar to values inferred for super-Earths.
Gas surface densities can be
depleted by more than five orders of magnitude relative to
the MMEN, and GCRs for $5 M_\oplus$ cores at the end
of 0.1 Myr will still remain above 0.01 (right panel, crosses).
If atmospheric growth is allowed to proceed
for 1 Myr (circles), then the nebula
can be depleted by more than eight orders of magnitude and
GCRs can stay above 0.01 (assuming
the fitted red dashed line can be extrapolated). 
Colored curves in the left panel denote various depleted
nebulae whose lifetimes are capped at $t = 1$ Myr; these
curves show that final GCRs $\geq$ 0.01 can be attained as long 
as the disk persists for $\gtrsim 8$ kyr,
which is reassuringly much shorter than typically modeled
disk clearing timescales.
The black curve is a sample evolution from the disfavored
scenario (a); here the planet barely escapes runaway in
an undepleted disk lasting $t = 10$ Myr. 
All models with nebular surface density $\Sigma_{\rm gas} < \Sigma_{\rm MMEN}/10^5$
are calculated with opacities extrapolated 
to low density
(see footnote \ref{footnote:opacity_extrapolation}).
}
\label{fig:rho_fudge}
\end{figure*}

Figure \ref{fig:rho_fudge} shows how this principle
is put into practice, underscoring just how little
ambient disk gas is needed to reproduce observed
GCRs. A $5M_\oplus$ core can reside
in a disk whose gas density is depleted relative to the 
MMEN by a factor as large as $10^7$ and still emerge with a
GCR of $\sim$2\% at the end of 1 Myr.
The rate at which a core
accretes gas hardly depends on the local
nebular density; the rate of accretion is
controlled by the rate of cooling, and cooling is regulated at
radiative-convective boundaries, which for the dusty atmospheres
featured in Figure \ref{fig:rho_fudge} are
isolated from the external nebula.
All other factors being equal, the final GCR
changes by a mere factor of 2 as the nebular depletion
factor runs from $10^2$ to $10^7$ (Figure \ref{fig:rho_fudge},
right panel). 
This insensitivity of the final GCR to nebular 
density applies only to disks that are depleted by 
at least factors of $\sim$10 relative to the MMEN;
in the left panel of Figure \ref{fig:rho_fudge}, we 
observe a larger-than-expected increase in the final GCR 
between the full $\Sigma_{\rm MMEN}$ case and 
more depleted cases. One reason for this jump
is the longer lifetime of a gas-rich
disk (10 Myr) vs.~that of a depleted disk (1 Myr).
Another reason is that for the gas-rich case,
the outermost layers of dusty atmospheres
are convective, whereas for most of the gas-poor
cases, they are radiative. Because radiative
layers have steep density gradients, they
more strongly decouple 
the rcb from the nebula at large.

Depleted inner disks of the kind envisioned here
bring to mind transitional
protoplanetary disks having central cavities or ``holes''
\citep[for a review, see][]{espaillat14}. 
A crucial element of our picture is that what
little gas pervades the inner disk (where super-Earths live)
is continuously replenished
by gas from the outer disk, over timescales not much
shorter than $\sim$1 Myr. It is both encouraging and surprising
that real-life transitional disks qualitatively fit this picture:
they have holes yet deliver gas onto their
central stars at rates comparable to those of disks
without holes. 
To resolve the paradox
of simultaneous transparency and accretion,
\citet{rosenfeld14}
posited that gas radial velocities are much larger inside the hole
than outside the hole, as gas surface density $\Sigma_{\rm gas}$
varies inversely
with inflow velocity at fixed disk accretion rate.
They entertained the possibility that gas radial velocities
could approach Keplerian velocities; found observational
evidence for free-fall velocities in the transitional disk
HD 142527; and speculated
that gravitational torques by massive companions
could drive these large velocities
\citep[see also][]{casassus15}.
We term this idea ``dynamic accretion,''
as distinct from the traditional and slower ``viscous
accretion.'' 
Technically the optically thin cavities
of transitional disks conflict with
our model for nebular accretion of atmospheres,
which assumes the disk to be optically thick.
We defer consideration
of how core accretion works in optically thin disks to a future
study.

\begin{figure*}[!tbh]
\centering
\includegraphics[width=\textwidth]{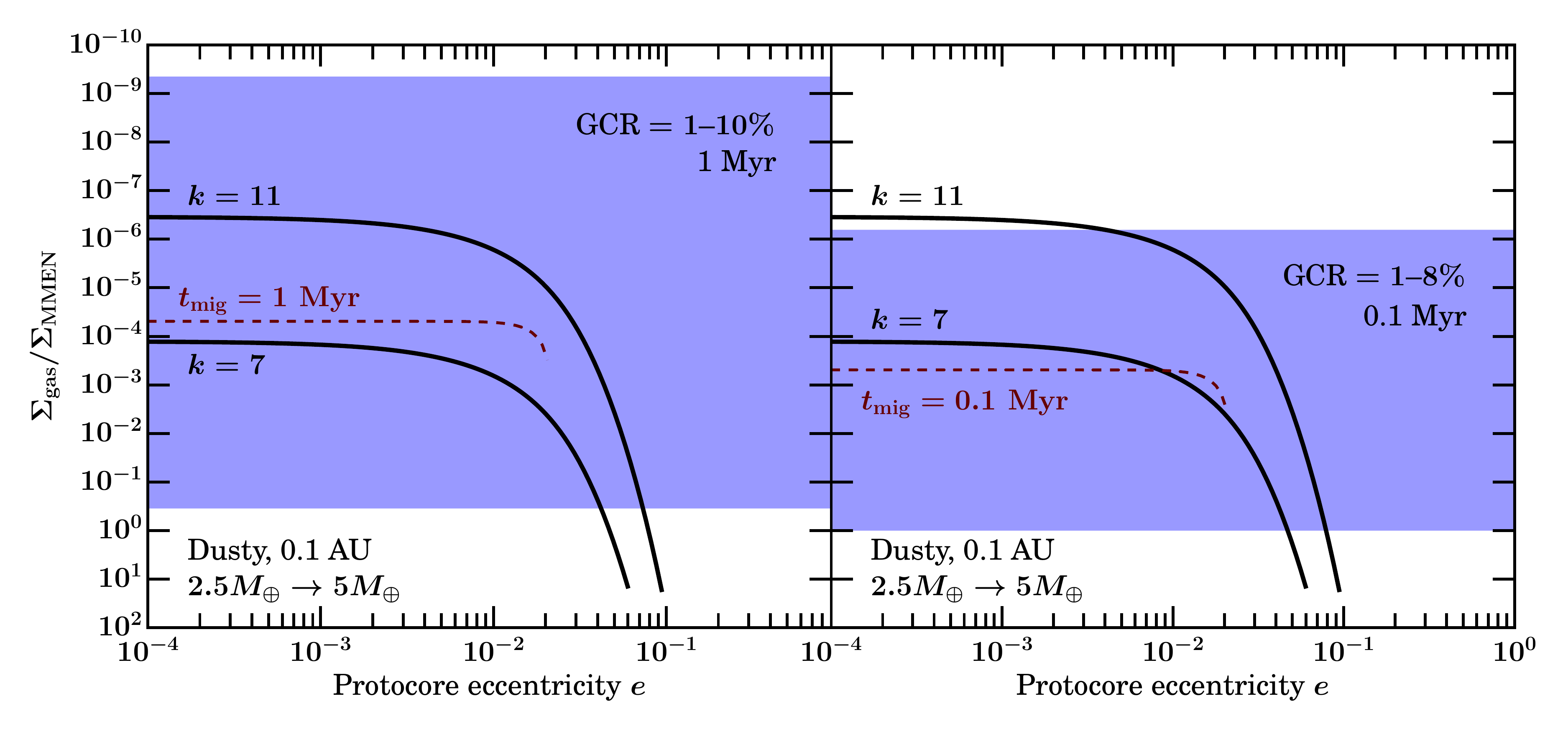}
\caption{\label{fig:minZ_e}
Super-Earth formation in a gas-poor nebula (scenario b)
accommodates a wide variety of disk environments and can
preserve planets against orbital migration.
Solid curves trace the amount of disk gas depletion
required for an ensemble of 2.5$M_\oplus$
protocores to ``viscously stir" one another onto crossing
orbits against gas dynamical friction, thereby merging into
5$M_\oplus$ cores. Each curve is calculated
from the condition $t_{\rm damp} = 0.1 t_X$
evaluated at $a = 0.1$ AU, and terminates 
at $\max e=kR_{\rm mH}/2a$, the orbit-crossing value. 
Orbital spacings $k$ (in units of mutual Hill radii $R_{\rm mH}$)
of 7 and 11 bracket possible protocore spacings 
as estimated from {\it Kepler} data (see text). 
Blue bars denote the gas depletion factors 
for which a 5$M_\oplus$ core at 0.1 AU can accrete a
dusty atmosphere having GCR = 0.01 -- 0.1 
within 1 Myr (left panel) and 0.1 Myr (right panel);
the time limit of 0.1 Myr is inspired by the
``photoevaporation-starved'' models
of \citet{owen11} in which the disk inside
$\sim$1 AU drains away on that e-folding time.
Note that within 0.1 Myr, 5$M_\oplus$ cores 
accrete up to only 8\% GCR in a full MMEN (right
panel).
That the blue bars overlap the $k$-curves
means that the constraints imposed by planetary GCRs and by
protocore mergers can be simultaneously satisfied;
indeed there are orders of magnitude of play
in the blue bars, reflecting the insensitivity
of atmospheric growth rates to the ambient nebular
density (see also Figure \ref{fig:rho_fudge}). 
Red dashed lines mark the gas surface densities 
for which
the Type I migration timescale $t_{\rm mig}$ \citep[equation 31 of][]{papaloizou00}
equals the assumed gas disk lifetime, evaluated for a
5$M_\oplus$ core at 0.1 AU.
For the most part, the constraints from protocore mergers
($k$-curves) sit safely above the 
migration constraint, indicating that migration may be negligible for most 
super-Earth systems.
}
\end{figure*}

In the left panel of Figure \ref{fig:minZ_e}, we indicate
with a colored bar the nebular gas densities that can
generate $0.01 \leq$ GCR $\leq 0.1$ within 1 Myr for
$5 M_\oplus$ cores having dusty atmospheres at 0.1 AU.
The range of gas depletion factors
$\Sigma_{\rm gas}/\Sigma_{\rm MMEN}$ 
is enormous, spanning nine orders of 
magnitude.\footnote{\label{footnote:opacity_extrapolation}
Nebulae that are depleted by more than five orders 
of magnitude with respect to the MMEN fall off the opacity table by
\citet{ferguson05} for dusty envelopes.
For these low-density nebulae, 
we extrapolate the opacity table by fitting a 
power law $\kappa \propto \rho^{\alpha} T^{\beta}$ 
for $-11 \leq \log\rho\,(\rhocgs) \leq -9$ and 
$2.7 \leq \log T ({\rm K}) \leq 3$.}
There is complementary flexibility in the accretion
time, which need not be 1 Myr; e.g., Figure \ref{fig:rho_fudge}
shows that a mere 8 kyr is required to attain
GCR = 0.01 if $\Sigma_{\rm gas} = \Sigma_{\rm MMEN}/100$.
Thus there is no need to fine-tune either disk properties
or the accretion time in scenario (b);
this conclusion will be supported further by calculations
below.

Whatever drives dynamic accretion (e.g., a massive
perturber, or disk self-gravity) may actually
de-stabilize the orbits of super-Earths.
An alternate, less dramatic way to generate inner disk holes
containing accreting gas is through ``photoevaporation-starved
accretion'' \citep{clarke01,drake09,owen11,owen12,alexander14}.
In this scenario, the disk interior to $\sim$1 AU becomes
starved when the disk outside $\sim$1 AU is blown out by a
photoevaporative wind before it can diffuse in.
A gap is formed near $\sim$1 AU, and
the starved interior disk has no choice
but to viscously drain onto the star and be
severely depleted. Super-Earth cores at $\sim$0.1 AU are
fed by gas that viscously diffuses from $\sim$1 AU,
but only for as long as it takes gas to diffuse
from that distance. This diffusion
time is short, on the order of $\sim$0.1 Myr
\citep[see, e.g., Figure 9 of][]{owen11}.
A time limit of 0.1
Myr for nebular accretion of planetary atmospheres
restricts the range of allowed gas depletion factors to those
shown in the colored bar of the right panel of Figure \ref{fig:minZ_e}.
The range of disk conditions compatible with GCR $\geq 0.01$
is narrower for an accretion time of 0.1 Myr vs.~1 Myr
(compare right vs.~left panels). But the range
of allowed disk gas densities
in the more time-limited case --- covering more
than six orders of magnitude --- still impresses.
Again, there is no indication of a fine-tuning problem. 

\begin{figure*}
\centering
\includegraphics[width=\textwidth]{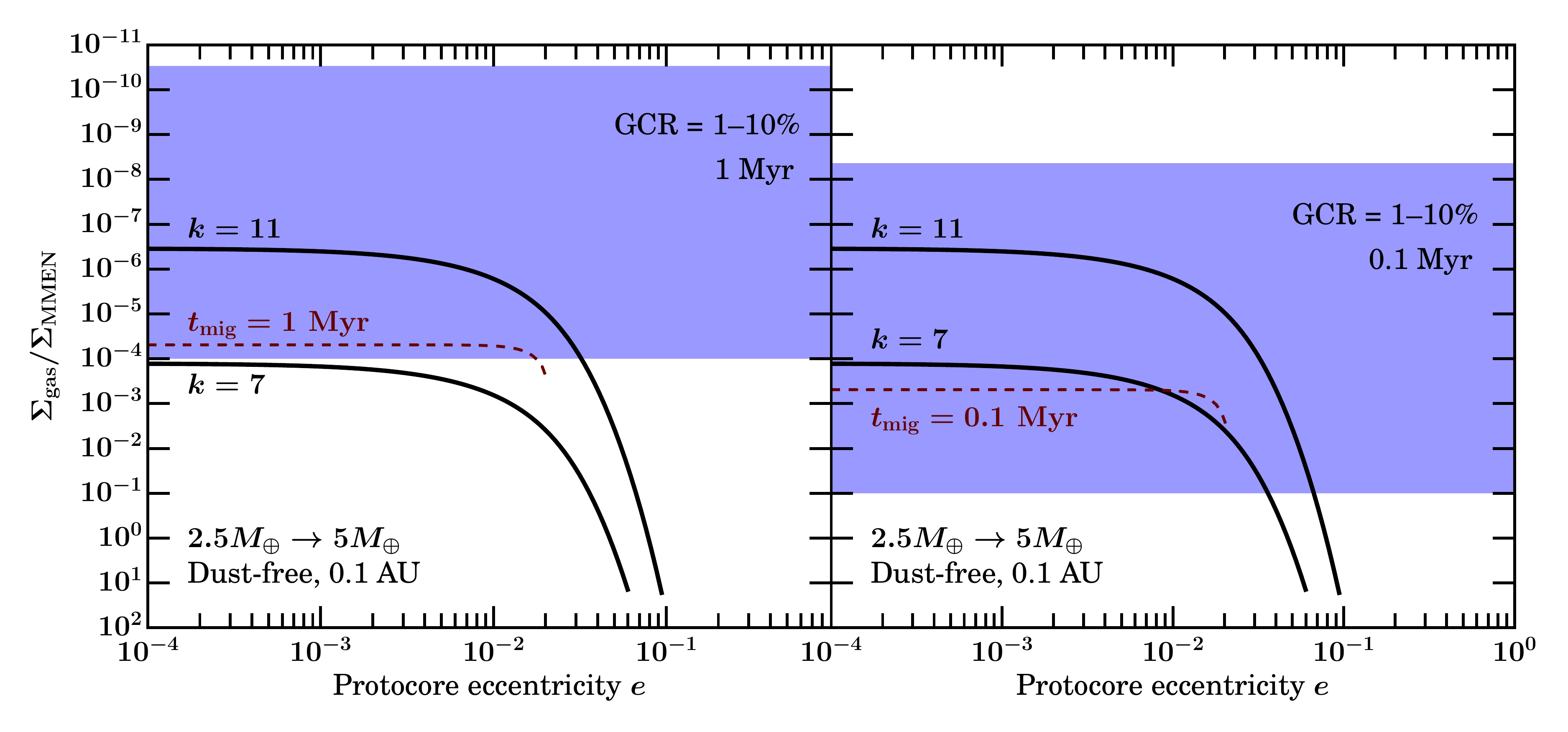}
\caption{\label{fig:maxZ_e_dustfree}Same as Figure \ref{fig:minZ_e} 
but for dust-free envelopes. The lower limits of colored 
bars correspond to nebular surface densities $\Sigma_{\rm gas}$ 
for which $5M_\oplus$ cores can emerge with GCR=10\% within 
1 Myr (left panel) and 0.1 Myr (right panel). 
The upper limits of colored bars correspond to 
gas depletion factors consistent with GCR=1\%, and are determined 
by extrapolating a power-law fit to GCR vs.~$\Sigma_{\rm gas}$ (such extrapolation
is necessary 
since neither our tabulated dust-free opacities nor analytic fits to those opacities
extend
to the lowest densities relevant here).
Gas depletion factors consistent with super-Earth 
GCRs span about 7--8 orders of magnitude, reflecting (as was the
case for the dusty envelopes of Figure \ref{fig:minZ_e})
the insensitivity of the cooling rates of envelopes to 
ambient gas densities.}
\end{figure*}

Figure \ref{fig:maxZ_e_dustfree}
is the same as Figure \ref{fig:minZ_e},
except that the colored bars are computed for
dust-free rather than dusty atmospheres.
Dust-free envelopes cool and accrete
faster than dusty envelopes and therefore
require more severe depletion
of the ambient gas disk to reach
a given GCR at a given time.
Nevertheless, dust-free envelopes
exhibit the same robustness
and tell the same story
as their dusty counterparts:
surface densities $\Sigma_{\rm gas}$
can range across 7--8 orders of magnitude,
and super-Earth cores can coagulate at
$\sim$0.1 AU and accrete dust-free atmospheres
having GCR $\sim$ 1--10\% within 0.1--1 Myr.

In Section \ref{ssec:deplete_merge} below, we 
compare the gas depletion factors computed here to those
needed to coagulate rocky cores against gas dynamical friction.
We also assess whether
super-Earth cores migrate 
in depleted gas disks (preview: they don't).

\subsection{Gas Depletion Factors Required to Merge Protocores}
\label{ssec:deplete_merge}

Unless 2--20$M_\oplus$
cores are born as isolation masses from disks with
high solid surface density (e.g., \citealt{dawson15}),
they form instead from collisions of smaller protocores
(a.k.a.~oligarchs).
Whether the target values of $\Sigma_{\rm gas}/\Sigma_{\rm MMEN}$ 
estimated in Section \ref{ssec:gas-depletion-factors}
are low enough 
to promote orbital instability
and mergers of protocores
depends on protocore
orbital spacings and eccentricities 
(and inclinations; 
Dawson et al., in preparation).
Mergers widen orbital 
separations, significantly 
increasing orbit-crossing 
timescales for subsequent mergers 
(e.g., \citealt{chambers96},~\citealt{zhou07}). 
Consequently, the time it takes a set of
isolation masses to merge into a system 
of super-Earth cores is largely determined by 
the final doubling: the transformation of
1--10$M_\oplus$ protocores into 2--20$M_\oplus$ cores.

We determine how depleted 
ambient gas disks must be
to effect the final doubling of protocores into full-fledged 
super-Earth cores. 
Formally, we solve for the conditions 
under which the eccentricity damping 
timescale $t_{\rm damp}$ exceeds
$\sim$0.1 of the orbit crossing timescale 
$t_X$~(the factor of 0.1 is empirically determined by \citealt{iwasaki01}; see also \citealt{kominami02}).\footnote{For
this last doubling, 
protocores are so widely spaced
(the number of Hill-radius spacings $k>5$) that the 
particle-in-a-box approximation 
(e.g., \citealt{safronov72}, 
\citealt{goldreich04}) likely fails \citep{ford07}.}
\citet{zhou07} give the following empirical
fitting formulae for $t_X$ for an ensemble
of nine equal-mass protocores situated
at a mean semimajor axis $a$, with adjacent
pairs separated by $\Delta a$:
\begin{align}
h &= \frac{k}{2}\left(\frac{2M}{3M_\ast}\right)^{1/3} \nonumber \\
k &= \Delta a / R_{\rm mH} \nonumber \\
R_{\rm mH} &= \left(\frac{2M}{3M_\ast}\right)^{1/3}a \nonumber \\
\label{eq:tX}
A &= -2 + e/h - 0.27\log_{10}\left(\frac{M}{M_\ast}\right) \nonumber \\
B &= 18.7-16.8e/h + (1.1-1.2e/h)\log_{10}\left(\frac{M}{M_\ast}\right) \nonumber \\
t_X &= 10^{A+B\log_{10}(k/2.3)}~{\rm yrs}~\left(\frac{a}{{\rm AU}}\right)^{1.5}
\end{align}
where $k$ is the orbital spacing 
in mutual Hill radii $R_{\rm mH}$,
$e$ is the eccentricity of a protocore,
$M$ is the mass of a protocore, 
and $M_\ast$ is the mass of the host star.
\citet{zhou07} provide results only for
protocores situated near $a = 1$ AU;
we have added a factor of $(a/{\rm AU})^{1.5}$
to equation (\ref{eq:tX}) so that we can scale
down to shorter orbital periods at $a \sim 0.1$ AU
where many {\it Kepler} planets reside. The true
scaling factor has not been carefully
measured and may differ slightly. 
Our calculations below set $a = 0.1$ AU
but clearly other formation distances are possible.

Eccentricity damping timescales 
are typically quoted for 
planets with small eccentricities $e \ll c_{\rm s}/v_{\rm K}$,
where $c_{\rm s}$ is the gas sound speed 
and $v_{\rm K}$ is the Keplerian velocity.
\citet{papaloizou00} find that 
when $e > c_{\rm s}/v_{\rm K}$,
damping becomes less efficient and
reduces to the form of classical 
dynamical friction ($\dot{e} \propto -1/e^2$).
We modify the eccentricity damping 
timescale of \citet{kominami02} 
to account for high eccentricity as follows:
\begin{align}
\label{eq:tdamp}
t_{\rm damp} = \frac{e}{|\dot{e}|} &= 0.5~{\rm yrs}~\left(\frac{T}{1000\,{\rm K}}\right)^{1.5} \left(\frac{6\times 10^{-6}\rhocgs}{\rho}\right) \nonumber \\ 
&\left(\frac{M_\oplus}{M}\right) \left[1+\frac{1}{4}\left(\frac{e}{c_s/v_{\rm K}}\right)^3\right] 
\end{align}
where $T$ is the gas temperature,
$\rho$ is the gas density,
and the factor in square brackets that
depends on $e$ is taken from equation (32)
of \citet{papaloizou00}. 
The volumetric density $\rho$ relates to the surface
density via $\Sigma_{\rm gas} = 
\sqrt{2\pi}\rho c_s a /v_{\rm K}$.

Values of $\Sigma_{\rm gas}/\Sigma_{\rm MMEN}$ 
implied by the 
condition $t_{\rm damp} > 0.1 \times t_X$ 
are shown as solid curves in Figure \ref{fig:minZ_e}.
Our plotted $k$-values of 7 and 11 were chosen as follows.
Observed {\it Kepler} multi-planet systems
(with multiplicity $\geq$ 5) 
have a present-day mean orbital spacing of
$k \simeq 14 \pm 3.4$  
(taken from the ``intrinsic" $k$ distribution of \citealt{pu15}; see their Figure 7).
Combining this range
with the assumption that
protocores maintain uniform masses and
spacings as they merge,
we estimate that just prior to the last doubling,
$k \sim (10.6$--$17.4) \times 2^{-2/3} \sim 6.7$--$11$ 
(the absolute spacing decreases by a factor of 2,
and protocores are half as massive as final cores so 
$R_{\rm mH}$ decreases by $2^{1/3}$). 

The smaller the protocore eccentricities $e$,
the smaller $\Sigma_{\rm gas}$ 
must be to effect mergers.
The target gas depletion factors
derived in Section \ref{ssec:gas-depletion-factors}
from GCR considerations are overlaid as colored
bars in Figure \ref{fig:minZ_e}.
For the most part,
the $k$-curves intersect the colored bars.
In other words,
the constraints imposed by super-Earth atmospheric
mass fractions (colored bars)
and by super-Earth core formation ($k$-curves)
can be simultaneously
satisfied for a wide range of 
protocore orbital spacings and eccentricities,
and a huge variety of
gas-poor inner disks (fed by gas-rich outer disks),
all over timescales of 0.1--1 Myr.

More precise determinations
of the gas depletion factors and 
the orbital properties of protocores 
required for core assembly --- i.e.,
better $k$-curves --- can be obtained
from $N$-body simulations that include gas damping and
that are tailored for close-in super-Earths.
\citet{kominami02}
and \citet{ogihara07} carried out similar calculations,
but for parameters specific to solar system terrestrial
planets.

We close this section
by noting that we expect eccentricities
of fully coagulated super-Earth cores 
to be damped to varying degrees
by residual gas \citep[see, e.g.,][]{agnor02}.
Just after their last doubling,
cores should have eccentricities of
order $e \sim 1/2 \times kR_{\rm mH}/2a \sim 0.03$--0.05
(the final $e$ should be the orbit-crossing value
just prior to the last doubling --- these are the values
at which the $k$-curves
in Figure \ref{fig:minZ_e} terminate --- multiplied
by 1/2 to account for momentum conservation
in inelastic collisions).
These eccentricities
can be damped by whatever gas remains:
we insert into equation
(\ref{eq:tdamp}) these $e$'s, together with
the depleted densities indicated by the
$k$-curves in Figure \ref{fig:minZ_e}
($\rho/\rho_{\rm MMEN} \sim 10^{-4}$ for $k=7$,
and $4\times 10^{-7}$ for $k=11$).
For $k=7$, we find that $5 M_\oplus$ cores at 0.1 AU
starting with $e \sim 0.03$
have an eccentricity
damping timescale $t_{\rm damp} \sim 0.002$ Myr; this
is much shorter than either the orbit crossing time
$t_X \sim 0.2$ Myr or the depletion time of a gas-poor disk
$t_{\rm disk} \sim 0.1$--1 Myr,
indicating that super-Earths created
under these conditions should have their orbits circularized.
For $k=11$, $t_{\rm damp} \sim 1$ Myr and
$t_X \sim 27$ Myr.
Because $t_{\rm damp}$ is now comparable
to the upper limit of $t_{\rm disk}$,
super-Earths created under these conditions
may have their starting eccentricities of $\sim$0.05
decreased by at most a factor of order unity.
Putting the results of $k=7$ and $k=11$ together,
we argue that super-Earth eccentricities today
should range anywhere from 0
to 0.05. These crude considerations align with observations:
from the distribution of transit durations of {\it Kepler}
planets, \citet{fabrycky14} find that rms inclinations
are $\sim$0.03 rad, and argue that rms eccentricities
should be similar.\footnote{More direct
constraints on eccentricities come from
statistical analyses of super-Earth
transit-timing variations (TTVs).
These constraints --- namely, rms eccentricities
of 0.018 \citep{hadden14} --- are relevant for systems near
mean-motion resonance; as such they may not be relevant
for run-of-the-mill super-Earths which are not resonant.}

\section{Super-Puffs as Migrated Dust-Free Worlds}
\label{sec:superpuff}

{\it Kepler} has discovered a relatively rare population of
especially large ($R \gtrsim 4 R_\oplus$)
and low mass ($M \lesssim 6 M_\oplus$)
planets. We refer to them as 
``super-puffs.''
Super-puffs are the exception, not the rule:
planets with $R > 4 R_\oplus$ are more than
an order-of-magnitude more rare at orbital
periods $< 50$ days than planets with $R < 4 R_\oplus$
(Fressin et al. 2013).\footnote{Super-puffs are a subset of what some call
``sub-Saturns''.} Examples include 
Kepler-51b \citep{steffen13}, a 
2.1$^{+1.5}_{-0.8}M_\oplus$ planet of size
7.1$\pm 0.3 R_\oplus$ orbiting its 
host star at $0.2$ AU \citep{masuda14}.
Another example is Kepler-79d \citep{steffen10}, 
a 6.0$^{+2.1}_{-1.6}M_\oplus$ planet of size
7.16$^{+0.13}_{-0.16}R_\oplus$ orbiting 
its host star at $0.3$ AU \citep{jontof-hutter14}.
The inferred GCRs of these two planets 
are 7--28\% and 33--40\%, respectively
\citep{lopez14}.

Can we fit a few super-puffs into our
theory of in-situ gas accretion from depleted nebulae?
Our numerical model 
gives ${\rm GCR} \sim 1\%$ for a $2M_\oplus$ core
(dusty atmosphere; 0.1 AU; depleted $\Sigma_{\rm MMEN}/200$
nebula lasting $t = 1$ Myr). Unfortunately, this GCR is much too low
compared to that of Kepler-51b.
Others have noted similar difficulties with
$R \gtrsim 4R_{\oplus}$ and $M \lesssim 5 M_\oplus$
planets accreting enough nebular gas in-situ at $\sim$0.1 AU
\citep{ikoma12, inamdar15}.

Nebular accretion can still explain
super-puffs (what other option is there?) --- perhaps
not at $\sim$0.1 AU where they are
currently found,
but rather at distances beyond $\sim$1 AU --- and with
the additional proviso that atmospheres be dust-free.
At larger orbital distances,
dust-free envelopes cool and accrete faster
than do dusty envelopes,
not just because of their reduced opacity
(Figure \ref{fig:dust-free}; see also 
\citealt{piso15}),
but also because dust-free atmospheres
are responsive to the colder temperatures
and lower densities characterizing the outer disk.\footnote{
Atmospheres can also be enriched with water beyond the
nebular ice line.
Water-rich atmospheres not only
have high mean molecular weight, but can also
have their density gradients steepened by the
thermostating effects of sublimating ice \citep{hori11,venturini15}.
More isothermal and higher $\mu$ atmospheres are more susceptible
to gravitational collapse onto low-mass cores on Myr timescales;
\citet{stevenson84} theorized that
sub-Earth-mass cores could rapidly nucleate gas giants
in icy nebulae, dubbing such planets ``superganymedean puffballs.'' 
The amount of water enrichment required 
to significantly accelerate gas accretion is large:
$Z \gtrsim 0.6$ (e.g., Figure 1 of \citealt{venturini15}).
But too large a $Z$ would render these planets
``water worlds'' or ``steam worlds,'' with
radii too small to match those of super-puffs
(e.g., Figure 7 of \citealt{lopez14}). Because
$Z$ would need to be fine-tuned to avoid
water worlds but still produce super-puffs, we
disfavor the idea that super-puffs accreted
their thick envelopes by virtue of their water content;
at the same time, we do expect super-puffs to have more water
than closer-in super-Earths.\label{foot:water}}
The radiative-convective boundaries of dust-free atmospheres
have temperatures approximately equal to the ambient
disk temperature. Because disk temperature decreases
with increasing orbital distance,
and because dust-free opacities diminish
with decreasing temperature,
dust-free atmospheres become optically thinner
farther away from the host star. It follows that more
distant, cooler, dust-free planets 
cool more easily and grow faster (Paper II).

Appealing to nebular accretion
beyond $\sim$1 AU requires that we 
invoke orbital migration to transport
super-puffs to their current observed locations
at $\sim$0.1 AU. 
Convergent migration of multiple
planets can lead to capture
into mean-motion resonances.
In this regard
we find it intriguing that all three
members of the Kepler-51 system (2 confirmed + 1
candidate; all are puffy, having
$R > 4 R_\oplus$) are situated near
a 1:2:3 resonance \citep{masuda14}.
The Kepler-79 system (3 super-Earths
+ 1 super-puff) may also be linked by
a 1:2:4:6 resonant chain \citep{jontof-hutter14}.

Type I migration requires gas. Timing super-puff migration
with disk dispersal --- in particular, the severe dispersal
required to form super-Earths (Section \ref{sec:sceB}) ---
may be tricky. It helps that super-puff cores
have less mass than super-Earth cores and can therefore
coagulate more readily in the face of
gas dynamical friction (cf.~Section \ref{ssec:deplete_merge}).
The right sequence of events might play out as follows:
super-puff cores coagulate --- perhaps as isolation masses --- 
and accrete their gas envelopes in a relatively
gas-rich disk outside 1 AU; they
migrate inward, possibly at the same time as they are amassing
atmospheres; they
get caught into resonance
with interior planets (either super-puffs or super-Earths);
and are finally parked inside 1 AU
as the gas disk --- now in its transitional phase --- clears
from the inside out
(i.e., as the central gas cavity grows from small radius
to large radius; cf.~\citealt{alexander14}; \citealt{chiang07}).
The rarity of close-in super-puffs might reflect
the difficulty in orchestrating this sequence
with favorable disk parameters.

\begin{figure}[!tbh]
\centering
\includegraphics[width=0.5\textwidth]{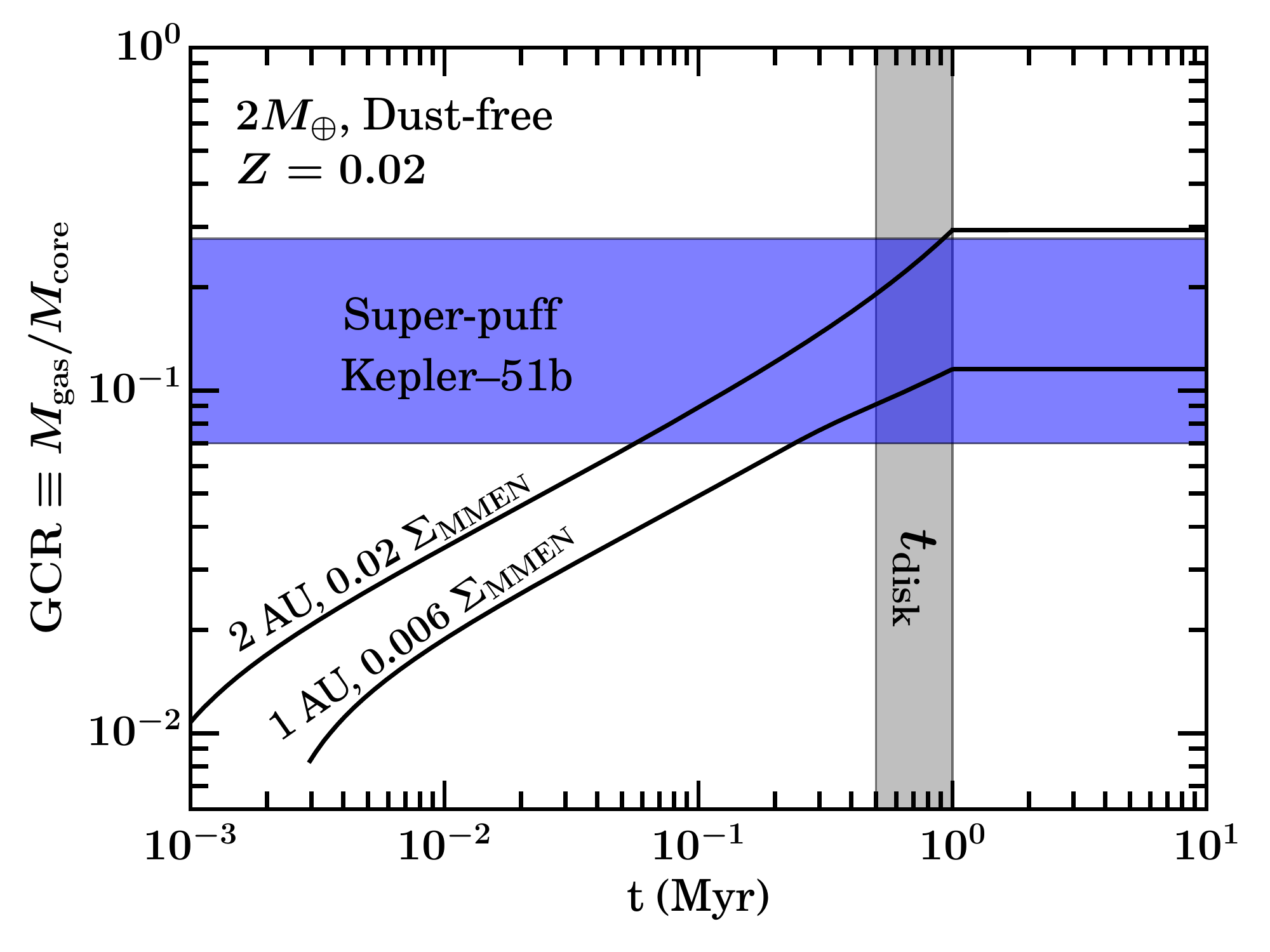}
\caption{\label{fig:gcr_M2_df}
Birth of a dust-free super-puff at 1--2 AU.
The blue region delineates the range of possible envelope fractions 
for 2$M_\oplus$ super-puff Kepler-51b, as calculated by
\citet{lopez14}, 
and the vertical gray bar denotes the timescales --- 0.1--1 Myr ---
over which the gas-poor disk finally clears.
(the clearing timescale $t_{\rm disk}$ for a gas-poor disk is 
$\lesssim$10\% that of a gas-rich disk).
Solid curves correspond to 2$M_\oplus$ cores 
embedded in nebulae whose surface densities (annotated on the plot)
are chosen to give a Type I migration timescale
$a/\dot{a} = 1$ Myr.
The relatively large GCRs characterizing super-puffs
can be achieved by having cores with dust-free atmospheres
form at larger orbital distances where nebular gas is cooler
and less opaque.}
\end{figure}

Uncertainties in migration history notwithstanding,
Figure \ref{fig:gcr_M2_df} 
demonstrates that super-puffs can readily
acquire their thick envelopes at large distances $\gtrsim 1$ AU.
Placed in a disk whose gas density
is large enough to permit inward migration
over $\sim$1 Myr, a 2$M_\oplus$ core at 1--2 AU
can accrete a dust-free atmosphere
with GCR $\sim 10$--30\%, successfully
reproducing the properties of Kepler-51b.

\section{Summary and Outlook}
\label{sec:conclusion}

{\it Kepler} super-Earths have core masses
large enough to nucleate gas giants,
yet have atmospheres that are 
modest by mass \citep[e.g.,][]{lopez14,wolfgang15}.
In Paper I, we proposed two possible scenarios
for making super-Earths over Jupiters: either (a) they form in
gas-rich nebulae
sufficiently enriched in dust
that atmospheres cool and therefore accrete
only slowly; or (b) they form in gas-poor (but not gas-empty)
disks, with no constraint on dust content.
Having gained additional
insights into the physics of nebular
accretion from Paper II,
we have in the present work re-evaluated these two scenarios
to determine which is more plausible.

Although strong radial gradients in dust content
are expected in protoplanetary disks (see, e.g., Paper I),
no amount of dust can save super-Earth atmospheres
from going runaway. Dust does not just increase the opacity
$\kappa$; it also increases (when sublimated at
atmospheric depth) the mean molecular weight $\mu$,
and high-$\mu$ atmospheres are prone to gravitational collapse.
At metallicities $Z \gtrsim 0.2$, the deadly side effects
of higher $\mu$ overwhelm the benefits of higher
$\kappa$. No cure is available for $\sim$10$M_\oplus$
cores in gas-rich nebulae lasting $\sim$10 Myr;
they are fated to spawn gas giants.

This leaves us with the gas-poor 
environment of scenario (b). 
Gas-poor conditions prevail just before
the disk disappears completely; such conditions are short-lived ($\sim$0.1--1 Myr vs.~$\sim$10 Myr) which helps to
stave off runaway (see Figure \ref{fig:minZ_e}).
A depleted nebula is also motivated on independent
grounds by the need for gas to deplete before
cores can coagulate. As disk gas vacates (either by viscously
draining onto the star or by being blown off in a wind),
dynamical friction weakens to the point where
protocores can gravitationally stir each other
onto crossing orbits and merge to become super-Earths.
The degree of gas depletion required to enable
protocores to merge can be severe: 
gas surface densities of disks may need to be 
$10^4$--$10^7$ times lower than those of a 
conventional, solar-composition 
minimum-mass nebula 
for viscous stirring to overcome gas dynamical friction.
Such tiny amounts of gas
might be thought to be insufficient
to furnish planetary atmospheric mass fractions $> 1$\%.
This is fortunately
not at all the case --- gas accretion rates onto planets
are incredibly insensitive to ambient disk gas
densities (see also Paper II). This
feature of nebular accretion should dispel concerns
that disk conditions need to be fine-tuned
to form super-Earth atmospheres.
As long as some gas can be siphoned from the outer disk
to feed the inner disk for 0.1--1 Myr, 
rocky cores in the inner disk can successfully 
pick up envelopes of the required mass.
The picture we are led to --- super-Earths forming
in an inner disk that contains only wisps of gas leaked
from a gas-rich outer disk --- is reminiscent
of transitional protoplanetary
disks, whose central cavities are not entirely void but contain
inflowing gas. So little gas is tolerated during
the era of core assembly that super-Earths are
safe from wholesale orbital migration driven by gas disk
torques: super-Earths may have formed in situ.

Nebular accretion under scenario (b) can also
accommodate super-puffs ---
a rare class of 
large radius ($R \gtrsim 4R_\oplus$),
low mass ($M \lesssim 6M_\oplus$) planets --- but
only if they formed 
at large orbital distances and migrated inward, unlike
their more sedentary super-Earth brethren.
The outer disk is more conducive
to atmospheric accretion because
it is colder. Colder gas is less opaque
and helps dust-free atmospheres cool
and amass faster.
For super-puffs to migrate inward by
gas disk torques, they must 
form in a nebula that is not too depleted;
the formation of super-puff cores pre-dates
the coagulation of super-Earth
cores. Super-puff cores may have
emerged as isolation masses at large
stellocentric distances beyond the ice
line. We might expect to find water
absorption 
features in the transmission spectra of
super-puffs:
infalling icy planetesimals and the erosion
of water-rich cores can pollute super-puff
envelopes. Our proposal that super-puffs
form at distances $\gtrsim 1$ AU
and migrate inward to their current
locations also
leads us to expect that super-puffs may be preferentially
part of mean-motion resonant chains.
Kepler-51 and Kepler-79 are candidate 
examples (\citealt{masuda14}; \citealt{jontof-hutter14}).
In particular, as one moves outward along a resonant chain,
one should find increasingly puffy planets.

Migration of super-puffs (but not super-Earths)
may explain why planet masses
and by extension bulk densities
obtained from transit-timing variations (TTVs)
are systematically low compared to those inferred from
radial velocity surveys
\citep{jontof-hutter14,weiss14,steffen15a,wolfgang15-prob}. 
Tightly spaced multi-planet systems situated
near mean-motion resonances are the most amenable
to TTV analyses. It may therefore be that TTVs are strongest
for super-puffs as we expect to find such objects chained
to resonances. Planet formation simulations also show
that puffier planets are more tightly packed
(Dawson et al., in preparation), further amplifying
their TTVs and biasing TTV studies to select
for especially low density objects.

\subsection{Future Directions}
\label{ssec:future}

In this series of three papers, we have detailed
some of the factors relevant for how planets accrete
gas from their parent disks
and quantified their primordial
gas-to-core mass ratios (GCRs).
We close now by taking
a step back and looking at the broader canvas
of issues --- what outstanding theoretical
questions need to be answered to mature the theory
of core nucleation, and how the theory fares against
observations.

\subsubsection{3D Hydrodynamics}
\label{sssec:3dhydro}
Our 1D calculation assumes that accretion flows onto solid cores 
are spherically symmetric. Three-dimensional numerical 
studies show a more complicated flow geometry: cores 
tend to accrete gas along the poles 
and expel some fraction of that gas along the equator
\citep[e.g.,][]{d'angelo13,ormel15,fung15}. 
How much 3D hydrodynamics and anisotropies
impact gas accretion rates is unclear. 
Isothermal 3D simulations \citep{ormel15,fung15}
find no bound atmosphere; material continuously
cycles into and out of the Hill or Bondi sphere.
By contrast, 3D simulations that solve the 
full energy equation with radiative cooling 
\citep{d'angelo13} find not only bound atmospheres but also 
gas accretion rates that are comparable to those
calculated in 1D.
While the results of \citet{d'angelo13} 
support 1D studies like ours,
their calculations are performed at $\sim$5 AU where 
3D effects may be muted because
the disk scale height there is safely larger
than the planet radius.
Future 3D investigations should focus
on smaller orbital distances and isolate the effects
of thermodynamics: adiabatic simulations
should be compared against isothermal simulations
and finally against calculations including radiative cooling.

\begin{figure}[!tbh]
\centering
\includegraphics[width=0.5\textwidth]{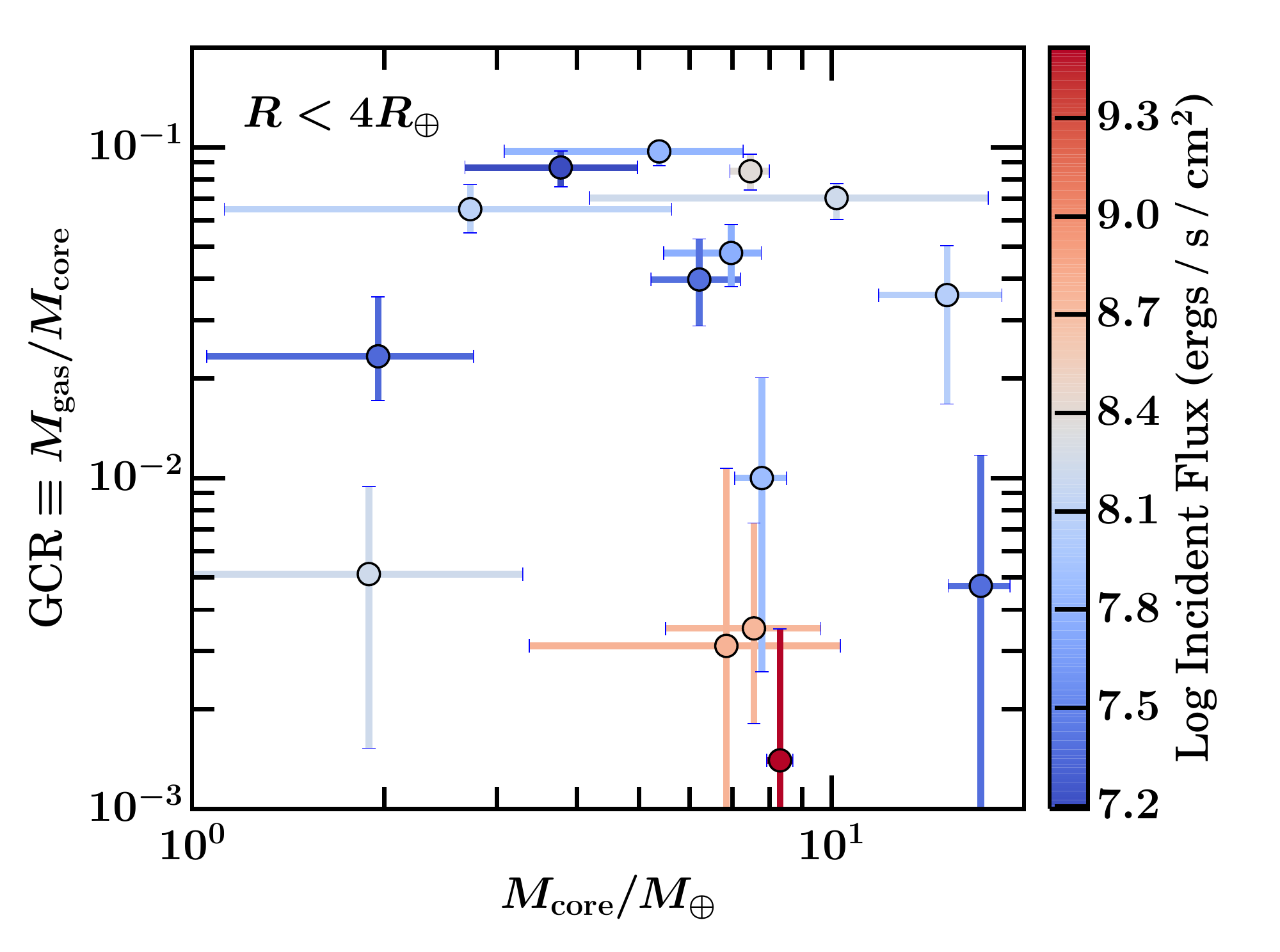}
\caption{\label{fig:LF15_gcr}
Gas-to-core mass ratios (GCRs) versus
core mass (data taken from Table 7 of
\citealt{lopez14}).\footnote{The published errors
in GCR (a.k.a.~envelope fraction)
for GJ 1214b, Kepler-11b, 55 Cnc e, Kepler-20c, 
Kepler-18b, Kepler-68b, and HD 97658b 
are inaccurate (Lopez \& Fortney, private communication).
For these objects, we used uncertainties re-calculated by the authors.}
The expectation from core accretion
theory is that more massive cores
should achieve greater gas envelope fractions ---
all other factors being equal.
From the large scatter in this figure,
it appears that all other factors
are not equal. Up to $\sim$6$M_\oplus$,
GCR increases with $M_{\rm core}$.
But at $M_{\rm core} \sim 6$--8$M_\oplus$,
GCR varies across two decades at nearly fixed
core mass; the variation appears correlated
with incident stellar flux (color shading)
and points to the influence of photoevaporative
erosion. At $M_{\rm core} > 10 M_\oplus$,
GCRs appear paradoxically to decrease
with increasing $M_{\rm core}$.
This unexpected behavior might arise because
core formation times lengthen with increasing
core mass; the most massive
cores might take so long to coagulate that
by the time they appear, their parent gas
disks have long departed.}
\end{figure}

\subsubsection{Correlation Between Core Mass and Envelope Mass}
\label{sssec:trends-core-mass-gcr}
More massive cores should accrete more massive atmospheres.
In Figure \ref{fig:LF15_gcr}, drawn from
models of planetary interiors by \citet{lopez14},
we examine whether this simple expectation is borne out.
It might be, for $M_{\rm core} \lesssim 6 M_\oplus$.
But for $M_{\rm core} \gtrsim 6 M_\oplus$,
the record of nebular accretion is hard to read.
A striking feature in Figure \ref{fig:LF15_gcr} is 
the large scatter in GCR, spanning two orders of magnitude,
within a narrow range of $M_{\rm core}\sim 6$--$8 M_\oplus$. 
By further color-coding these data by incident stellar flux,
we observe that smaller GCRs correlate with larger
irradiances, and deduce that evaporative winds powered by stellar irradiation
may have eroded GCRs \citep[e.g.,][]{lopez13,owen13}.
The data are scant and hardest to interpret 
for $M_{\rm core} > 10 M_\oplus$; here
the largest mass cores happen to be the least irradiated
yet have the smallest GCRs.
Perhaps core formation time is the ``hidden variable'':
if the most massive cores take longer than the gas disk
lifetime to coagulate, they will fail to acquire
massive atmospheres. 
Core assembly times can vary by orders of magnitude,
depending exponentially on the core mass and
the solid surface density of the parent disk \citep[e.g.,][]{dawson15}.

\subsubsection{Super-Earths vs.~Jupiters:
\\Trends with Orbital Distance and Host Stellar Type}
\label{sssec:super-earths-vs-jupiters}

Our expectation that puffier planets form more easily
at larger orbital distances may be tested 
in individual multi-planet systems. 
Among the planets with 
both radii and mass measurements \citep{masuda14,weiss14,jontof-hutter15},
we identify 5 out of 9 systems with more than 3 transiting planets 
that follow the predicted trend of decreasing bulk density
with increasing orbital period: Kepler-20~\citep{gautier12}, 
Kepler-37~\citep{marcy14}, Kepler-79~\citep{jontof-hutter14}, 
Kepler-102~\citep{marcy14}, and Kepler-138~\citep{jontof-hutter15}. 
Of the remaining 4 systems, 3 are consistent with 
the above density trend within errors.
The effects
of photoevaporative erosion \citep{owen13}
and observational bias (long periods select
for larger and presumably puffier planets in transit searches)
need to be subtracted off
to more cleanly test our theory of atmospheric accretion.

Around FGK stars probed by radial velocity surveys,
Jupiters are known to exist preferentially at large orbital
distances \citep[e.g.,][]{cumming08}. 
It is tempting to ascribe
the observed rapid rise in gas giant
occurrence rate at $\sim$1 AU 
to the water ice line in protoplanetary disks.
Just outside the water condensation
front, the disk's surface density in solids is boosted by ice
so that cores massive enough to undergo runaway gas accretion
coagulate more quickly, within the gas disk lifetime.
Water also accelerates nebular accretion of atmospheres,
by virtue of its high mean molecular weight and chemical potential
(see footnote \ref{foot:water}).

Whether the rise in the gas-giant frequency at $\sim$1 AU
coincides with a fall in the super-Earth frequency at that
same distance is unclear. Such a fall would signal the transformation
of super-Earth cores into Jupiters and would provide
strong support for the theory of core accretion.
In Paper I (section 5),
we pointed to some data for FGK stars
in \citet{fressin13} that hinted at such a fall 
\citep[see also][]{dong13,burke15}.
But the situation is too early to call.
It may be instead
that the occurrence rate of super-Earths is roughly
flat with orbital distance --- that super-Earths
are truly ubiquitous --- as
appears to be the case for M stars, as revealed by microlensing 
\citep[e.g.,][]{sumi10,gould10,cassan12,clanton14,clanton15}
and transit surveys \citep[][]{dressing15}.
These same studies indicate that M stars
are less likely to host gas giants than FGK stars
(see also \citealt{bonfils13}).

Perhaps the simplest picture we can paint with the data
and theory currently in hand is that most stars in the universe are
born with disks having solid surface densities that are neither
too ``low'' nor too ``high''.
The higher the solid surface density, the faster
the coagulation of rocky cores \citep{dawson15}.
Most disks have ``medium'' solid surface densities
in the sense that they spawn
super-Earth cores near the tail-end of the gas lifetime
(scenario b) --- neither so fast that cores are birthed 
into gas-rich environments to become Jupiters,
nor so slow that they fail to acquire any atmosphere at all
because the parent gas disk has completely dissipated.
We have seen in this paper
that this middle ground is enormous and presents no fine-tuning
problems.
A minority of FGK stars --- about
$\sim$10\% \citep{cumming08} --- 
have disk surface densities
just high enough to spawn Jupiters, and then only with the added boost
from the ice line at $\sim$1 AU. 
In this picture,
there should be a corresponding
drop in the occurrence rates of super-Earths at that distance
around FGK stars. As for M dwarf disks, we posit
that their surface densities are systematically lower than those of
their FGK counterparts
(see also \citealt{laughlin04}; \citealt{andrews13};
cf.~\citealt{kornet06}), so much so that they almost never
produce gas giants, but instead make super-Earths everywhere.
The unifying theme in this rough sketch is disk solid
surface density and its decisive impact on planet demographics.

\acknowledgments
Our high-$Z$ analysis was made possible by 
Jason Ferguson who generously calculated all the opacity tables for 
$\log T\, ({\rm K}) > 2.7$.
We thank Heather Knutson, Chris Ormel, James Owen,
and an anonymous referee for careful 
readings of our manuscript that led to substantive improvements, 
and Gennaro D'Angelo, Rebekah Dawson, 
Jonathan Fortney, Jeffrey Fung, Eric Lopez, Ruth Murray-Clay, 
Sergei Nayakshin, and Yanqin Wu
for helpful discussions. 
We are also grateful to
Daniel Fabrycky and Hilke Schlichting 
for motivating us to consider the formation of 
super-puffs; David Bennett, Christian Clanton,
and Scott Gaudi for educating us about
planet occurrence rates around M stars
as measured by microlensing surveys;
and Jeff Cuzzi and Jack Lissauer
for prompting us to revisit the 
assumption of dusty atmospheres.

EJL is supported in part by the Natural Sciences and 
Engineering Research Council of Canada under PGS D3 and 
the Berkeley Fellowship. EC acknowledges support from 
grants AST-0909210 and AST-1411954 awarded by the 
National Science Foundation, NASA Origins grant 
NNX13AI57G, and Hubble Space Telescope grant HST-AR-12823.001-A.

Numerical calculations were performed on the SAVIO 
computational cluster resource provided by the 
Berkeley Research Computing program at the 
University of California Berkeley, supported by 
the UC Chancellor, the UC Berkeley Vice Chancellor 
for Research, and Berkeley's Chief Information Officer.

\bibliography{followup}

\end{document}